\documentclass[twocolumn]{aastex631}

\DeclareUnicodeCharacter{2212}{-}

\usepackage{amsmath}
\usepackage{hyperref}

\usepackage{savesym}
\savesymbol{tablenum}
\usepackage{siunitx}
\restoresymbol{SIX}{tablenum}

\newcommand{\Msun}{M_{\odot}}
\newcommand{\Lsun}{L_{\odot}}

\newcommand{\kms}{\mbox{${\rm km~s^{-1}}$}}
\newcommand{\kmsMpc}{\mbox{km s$^{-1}$ Mpc$^{-1}$}}

\begin{document}


\title{The TRGB$-$SBF Project. II. \\ Resolving the Virgo Cluster with JWST} 


\author[0000-0002-5259-2314]{Gagandeep S. Anand}
\affiliation{Space Telescope Science Institute, 3700 San Martin Drive, Baltimore, MD 21218, USA}

\author[0000-0002-9291-1981]{R. Brent Tully}
\affiliation{Institute for Astronomy, University of Hawaii, 2680 Woodlawn Drive, Honolulu, HI 96822, USA}

\author[0000-0001-5487-2494]{Yotam Cohen}
\affiliation{Space Telescope Science Institute, 3700 San Martin Drive, Baltimore, MD 21218, USA}

\author[0000-0002-3234-8699]{Edward J. Shaya}
\affiliation{Astronomy Department, University of Maryland, College Park, MD 20743, USA}

\author[0000-0001-9110-3221]{Dmitry I. Makarov}
\affiliation{Special Astrophysical Observatory of the Russian Academy of Sciences, Nizhnij Arkhyz, Karachay-Cherkessia 369167, Russia}

\author[0000-0003-0736-7609]{Lidia N. Makarova}
\affiliation{Special Astrophysical Observatory of the Russian Academy of Sciences, Nizhnij Arkhyz, Karachay-Cherkessia 369167, Russia}

\author[0009-0004-4126-8924]{Maksim I. Chazov}
\affiliation{Special Astrophysical Observatory of the Russian Academy of Sciences, Nizhnij Arkhyz, Karachay-Cherkessia 369167, Russia}

\author[0000-0002-5213-3548]{John P. Blakeslee}
\affiliation{NSF's NOIRLab, 950 N Cherry Ave, Tucson, AZ 85719, USA}

\author[0000-0003-2072-384X]{Michele Cantiello}
\affiliation{INAF $–$ Astronomical Observatory of Abruzzo, Via Maggini, 64100, Teramo, Italy}

\author[0000-0001-8762-8906]{Joseph B. Jensen}
\affiliation{Department of Physics, Utah Valley University, 800 W. University Parkway, Orem, UT 84058, USA}

\author[0000-0002-5514-3354]{Ehsan Kourkchi}
\affiliation{Institute for Astronomy, University of Hawaii, 2680 Woodlawn Drive, Honolulu, HI 96822, USA}
\affiliation{Department of Physics, Utah Valley University, 800 W. University Parkway, Orem, UT 84058, USA}

\author[0000-0002-5577-7023]{Gabriella Raimondo}
\affiliation{INAF $–$ Astronomical Observatory of Abruzzo, Via Maggini, 64100, Teramo, Italy}


\begin{abstract}

The Virgo Cluster is the nearest substantial cluster of galaxies to the Milky Way and a cornerstone of the extragalactic distance scale. Here, we present JWST/NIRCam observations that simultaneously cover the cores and halos of ten galaxies in and around the Virgo Cluster and are designed to perform simultaneous measurements of the tip of the red giant branch (TRGB) and surface brightness fluctuations (SBF). Seven of the targets are within the Virgo Cluster and where we are able to resolve some of the cluster's substructure, while an additional three provide important constraints on Virgo infall. The seven galaxies within Virgo itself all have SBF measurements from the Advanced Camera for Surveys Virgo Cluster Survey (ACSVCS). After adjusting the ACSVCS measurements for the offset from our new JWST TRGB measurements, we determine a distance to the Virgo Cluster of d~$=$~16.17~$\pm$~0.25~(stat)~$\pm$~0.47~(sys)~Mpc. The work presented here is part of a larger program to develop a Population II distance scale through the TRGB and SBF that is completely independent of the prominent Cepheid + Type Ia supernova ladder. The TRGB distances to the galaxies presented here, when combined with future SBF measurements, will provide a crucial step forward for determining whether or not systematic errors can explain the well-known ``Hubble tension'' or if there is significant evidence for cracks in the $\Lambda$CDM model. 

\end{abstract}




\section{Introduction}\label{sec:intro}

The Hubble tension, a discrepancy between $\Lambda$CDM-based predictions and locally-determined values of the Hubble constant, persists \citep{2021arXiv210301183D}. The disagreement between the $\Lambda$CDM- predicted value ($H_{0}$ = 67.4 $\pm$ 0.5 \kmsMpc) using cosmic microwave background (CMB) observations from Planck \citep{2020A&A...641A...6P} and the ``gold-standard" distance ladder of Cepheids and Type Ia supernovae from the SH0ES team of $H_{0}$ = 73.17 $\pm$ 0.86 \kmsMpc \citep{2022ApJ...934L...7R, 2023JCAP...11..046M, 2024ApJ...973...30B} exceeds 5$\sigma$ in significance.

In recent years, the role of the tip of the red giant branch (TRGB; \citealt{1993ApJ...417..553L, 2009AJ....138..332J, 2016ApJ...832..210B, 2017AJ....154...51M,2019ApJ...880...52A}) has become more prominent with regards to the Hubble tension, due to its use as a replacement for Cepheids within the second-rung of the SH0ES distance ladder. The method relies on the standard candle nature of the brightest ascent of low-mass red giants. Notably, the Carnegie-Chicago Hubble Program (CCHP; \citealt{2016ApJ...832..210B}) has reported a value of the Hubble constant from TRGB and Type Ia supernovae of $H_{0}$ = 69.8 $\pm$ 0.8 (stat) $\pm$ 1.7 (sys) \kmsMpc \citep{2019ApJ...882...34F}, which lies almost directly between the Planck and SH0ES values, perhaps easing the tension. However, alternate reductions and analyses of the same \textit{HST} CCHP datasets provide values closer to the original SH0ES result \citep{2022ApJ...932...15A, 2023ApJ...954L..31S}. First results from JWST largely appear to only solidify the tension \citep{2023ApJ...956L..18R, 2024ApJ...962L..17R, 2024ApJ...966...89A, 2024ApJ...976..177L}, with some disagreement from other ongoing studies \citep{2024arXiv240803474L, 2024arXiv240806153F}, though these may simply be the result of limited sample sizes \citep{2024ApJ...977..120R}.

While such details are still being resolved within the literature, it is vital to note that the route proposed by the CCHP \textit{cannot} serve as an independent check on the SH0ES results, as they and SH0ES both make use of Type Ia supernovae as the final-rung of the distance ladder. If we as a community are to make the claim that the Hubble tension truly exists, and is not the result of systematic errors along the distance ladder, then we must provide an independent pathway than that given by the SH0ES team.

A promising alternative to Type Ia supernovae are Surface Brightness Fluctuations (SBF; \citealt{1988AJ.....96..807T, 2015ApJ...808...91J, 2023arXiv230703116C}), which which use the spatial fluctuations in surface brightness as a measure of distance. \cite{Blakeslee2021} used SBF distances measured by \citet[][]{2021ApJS..255...21J} from WFC3/IR observations of 63 elliptical galaxies out to 100~Mpc 
 to derive $H_{0}$ = 73.3 $\pm$ 0.7 (stat) $\pm$ 2.4 (sys) \kmsMpc. The calibration of their SBF distances was based on a mix of six Cepheid and two TRGB distances. They also noted that their systematic uncertainty could be reduced by using JWST to measure TRGB distances to a sizable sample of nearby, early-type galaxies.

It is within this context that we began the TRGB$-$SBF Project. The goal of this project, as also described by \cite{2024ApJ...973...83A}, is to link the distance scales obtained by TRGB and SBF through simultaneous observations with the same instrument (JWST/NIRCam) to reduce sources of systematic error, while also measuring SBF distances out into the Hubble flow for an eventual 1$\%$ determination of $H_{0}$ independent of Cepheids and Type Ia supernovae.

So far, our team has been awarded time with two JWST programs. The first is the Cycle 2 program GO–3055 \citep{2023jwst.prop.3055T}, from which we draw data for this paper. This program has observed fourteen nearby elliptical galaxies with NIRCam for purposes of tying the TRGB and SBF distance scales. The targets were drawn from nearby elliptical and lenticular galaxies expected to be nearer than 20~Mpc, with K-band intrinsic luminosities such that $\log_{10}(L_{K}) >11$. From 18 candidates, we rejected four based on color and dust properties which would indicate they are not well suited for high-precision SBF measurements \citep{2010ApJ...724..657B}.

A second program, JWST GO-5989 \citep{2024jwst.prop.5989J}, was accepted for Cycle 3, and will provide observations of $\sim$50 massive early-type galaxies in the Coma cluster for purposes of fully calibrating SBF age and metallicity effects in the JWST filter system. We will also require observations of elliptical galaxies out into the Hubble flow for the final rung of our proposed distance ladder. The end result would be a value of $H_{0}$ that is independent of both Cepheids \textit{and} Type Ia supernovae. More work will also be required to further develop the initial geometric calibration of the TRGB (see the discussion in Section 5).

At the heart of both of the TRGB and SBF distance indicators are red giant branch stars. The short-wavelength observations in $F090W$ and $F150W$ are used to construct color-magnitude diagrams to measure TRGB distances. As described in detail by \cite{2024ApJ...966...89A}, $F090W$ is an optimal choice for the primary filter due to the stability of the TRGB as a function of age and metallicity. While the TRGB is brighter in redder filters, the slope becomes difficult to ascertain \citep{2024ApJ...975..111H}, leading to increased uncertainties which would prohibit an eventual 1$\%$ measurement of $H_{0}$. These same filters can be used for optimal SBF measurements. Long-wavelength observations in $F277W$ and $F356W$ are obtained simultaneously due to the multiplexing capabilities of NIRCam, and will be useful for SBF stellar population studies \citep{2001MNRAS.320..193B} as well as additional science cases by the broader community, such as determining the recent star-formation histories of these galaxies \citep{2024ApJ...976...60M, 2024arXiv241009256L}.

In this paper, we present JWST/NIRCam data for seven galaxies within the Virgo Cluster, as well as three additional galaxies that are nearby, and provide constraints on Virgo infall and its zero-velocity surface. We provide distances to each of these galaxies, as well as a revised distance to the Virgo Cluster. Similar results for three galaxies in the Fornax Cluster were presented by Paper I in this series \citep{2024ApJ...973...83A}. Observations for one remaining galaxy, NGC~1549 in Dorado were executed later (in the Fall of 2024), and will be presented in a future work. The TRGB distances from this work will provide the foundation for the zero-point scaling of more distant SBF measurements into the Hubble flow, and a revised value of the Hubble constant.

\section{The Virgo Cluster}

Before we present our JWST observations, we should first address the question: What is the Virgo Cluster?
Its locally-dominant and complex nature has long been recognized; 14 of 34 objects now recognized as galaxies in the original Messier catalog \citep{MessierCatalog} project onto a $5^{\circ}$ radius patch of the sky in Virgo. However, the related structure is complicated.  \citet{deVaucouleurs1961} recognized that components that overlap in velocity but lie at different distances are juxtaposed and introduced some of the names of features still in use: the Virgo Southern Extension, Virgo\,W, and Virgo\,W$^\prime$ (although the composition of Virgo\,W$^{\prime}$ is different today).

Virgo\,W is particularly problematic because by itself it is as substantial as the Fornax Cluster. Lying at about twice the distance of the Virgo Cluster, it abuts Virgo to the southwest in equatorial coordinates, fortunately with only slight spatial overlap. Unsurprisingly, though, there are outliers in the Virgo\,W vicinity that contaminate Virgo Cluster galaxy samples, mainly in what is known as the M Cloud \citep{Ftaclas1984}. 
Members of these structures cannot be distinguished based on morphologies or velocities alone but are distant enough to be separated with Fundamental Plane \citep{1987ApJ...313...42D} or Tully-Fisher  \citep{1977A&A....54..661T} distances.

Nearer to Virgo, the disposition of a grouping at the location of de Vaucouleurs' Virgo\,W$^\prime$ has only become clarified with the relative accuracy of SBF distance measurements with Hubble Space Telescope \citep{2007ApJ...655..144M, Blakeslee2009}, and an even more extensive SBF distance study using  high-quality wide-field imaging from the Next Generation Virgo Survey \citep{Cantiello2024}.
Virgo\,W$^\prime$ lies $\sim\,$40\% farther than Virgo with almost the same mean velocity. 
A supernova event and accompanying Cepheid observations in one of its members, NGC\,4639, provide distances consistent with the SBF measurements \citep{2001ApJ...553...47F, Jha2007}.
A small number of additional galaxies at a similar distance are identified as Virgo\,W$^{\prime\prime}$ by \citet{Kourkchi2017} in their Fig.\,2.
\citet{Kourkchi2017} also identify the locations of several dwarf galaxies that have turned up with Hubble Space Telescope imaging to lie in the foreground at $\sim 10$~Mpc \citep{2014ApJ...782....4K}.  It is difficult to estimate the distances of such dwarfs based on morphology.  It can be anticipated that more small foreground systems will be found in front of the Virgo Cluster with deep imaging.

The boundaries of the Virgo Cluster are rather well defined in all directions except to the south, de Vaucouleurs' Southern Extension.
Whereas the velocity dispersion within the domain of the Virgo Cluster is large, there are groupings of locally coherent velocities in the Southern Extension \citep{Tully1984, Kourkchi2017}. 
It can be inferred that galaxies in the Southern Extension are falling with close to radial orbits toward the main Virgo Cluster.  These galaxies will very likely be accreted by the cluster.
Collectively, their motions can be modeled to give the mass of the Virgo Cluster of $6.3 \pm0.8 \times 10^{14} ~\Msun$ \citep{2017ApJ...850..207S, Kashibadze2020}.

The influx of infalling galaxies into the Virgo Cluster is substantial, as initially inventoried by \citet{Tully1984} and updated by \citet{Kourkchi2017}.
The integrated $K_s$-band light is roughly equal ($5\times 10^{12}~\Lsun$) in both the Virgo Cluster proper and the infalling domain.  From orbit modeling, the Virgo Cluster will grow 30\% by mass in a Hubble time \citep{2017ApJ...850..207S}.  
The infalling galaxies have lower mass-to-light ratios than average cluster members, reflective of the greater proportion of gas-rich systems actively forming stars outside the cluster.

Three galaxies in the current JWST program in proximity to the Virgo Cluster provide constraints.
NGC\,4697 and M105 each lie near the transition surface between cluster infall and cosmic expansion. 
NGC\,4636 is particularly interesting as the dominant member of a group that is only 10\textdegree\ from the Virgo core galaxy M87.  As discussed in section 4.5, this significant group will enter the cluster within a Gyr.
It cannot be doubted that a considerable number of galaxies within the Virgo cluster are recent arrivals.

\begin{figure}
\epsscale{1.23}
\plotone{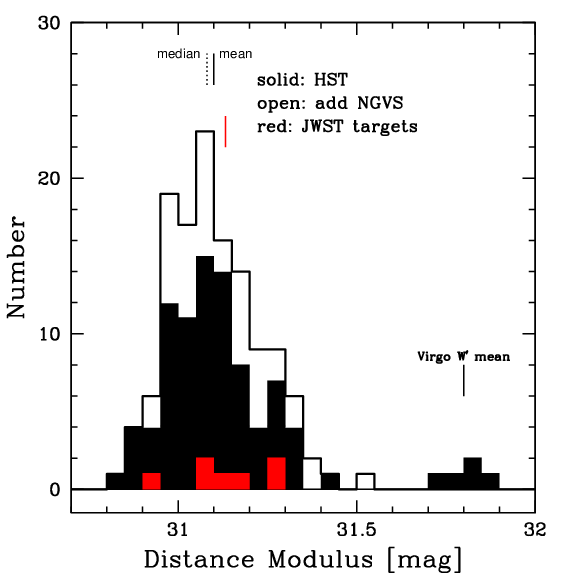}
\caption{Histogram of distances to members of the Virgo Cluster and background Virgo W$^{\prime}$ Cluster from the HST ACS and NGVS SBF surveys. The filled histogram is from ACS alone, and the open histogram is from the sum of the two studies.  The filled red histogram shows the values in these surveys associated with the JWST targets discussed in this paper.  The median and mean distance modulus values for the Virgo ensemble are 31.10 and 31.08 mag, respectively. The mean value for the seven JWST cases is 31.13 mag. The median distance modulus value for Virgo\,W$^{\prime}$ is 31.80 mag.
}
\label{fig:histDM}
\end{figure}


For the present discussion, the issue is the degree to which the modest number of JWST measurements to be discussed defines the cluster distance.
The major constituents of the cluster have been revealed by two particularly important surveys: the photographic catalog of galaxies in the Virgo Cluster area obtained with the du~Pont 2.5m telescope at Las Campanas Observatory \citep{1985AJ.....90.1681B} and the Next Generation Virgo Cluster Survey (NGVS) carried out with a wide-field CCD camera at the Canada-France-Hawaii Telescope on Maunakea, Hawaii \citep{Ferrarese2012}.
Given the importance of the Virgo Cluster, there have been complementary surveys across the electromagnetic spectrum, ranging from the radio at HI with the Arecibo Telescope \citep{Haynes2011, Haynes2018} to X-ray with ROSAT \citep{1994Natur.368..828B}. 
There is a description of the principal substructures in the presentation of the GALEX Ultraviolet Virgo Cluster Survey \citep{Boselli2014}.  
That article draws attention to components called A, B, and C, around M87, M49, and M60 respectively, in addition to a region favored by galaxies with low systemic velocities, the Low Velocity Cloud (LVC).  
Density, morphology, and velocity patterns across the cluster are strong inferences of continued accretion into the cluster.
\citet{Binggeli1993} speculate that there is a distinct sub-clump around M86 with relative negative velocities with respect to M87 in the core of the cluster.

\begin{figure*}
\epsscale{1.15}
\plotone{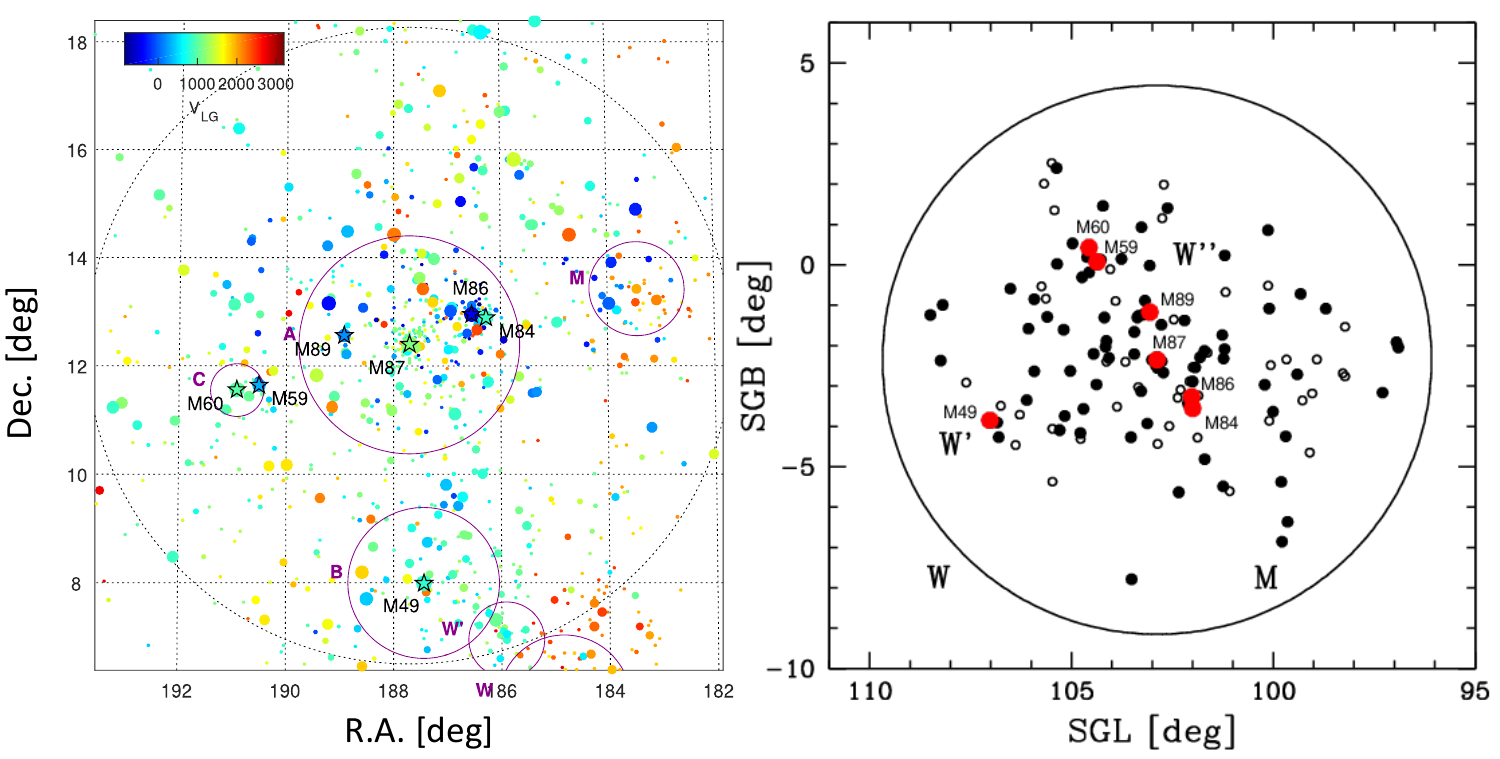}
\caption{\textbf{Left:} Galaxies with Local Group frame velocities $ < 3200 $~\kms\ are projected onto the Virgo Cluster in equatorial coordinates.  Circles locate the cluster substructures A, B and C and the background W$^{\prime}$ and M structures.
\textbf{Right:} Distribution of 128 galaxies from the combined HST ACS and NGVS samples in supergalactic coordinates.  Solid circles identify the HST cases and open circles identify remaining NGVS cases.  The 7 targets of the JWST program are located with red symbols and named. The letters W and M are at the positions of structures at twice the Virgo distance, and W$^\prime$ and W$^{\prime\prime}$ are the locations of structures 40\% further than Virgo.  The outer circle lies at radius $6.8^\circ$ about M87.}
\label{fig:allsbf}
\end{figure*}

Surface brightness fluctuation (SBF) \citep{1988AJ.....96..807T} distance  measurements have been made with sufficient accuracy to resolve the depth of the Virgo Cluster and reject background interlopers.
Two extensive and coherent studies are important for the present discussion.
The first is a product of the ACS Virgo Cluster Survey (ACSVCS) with HST \citep{Cote2004} that gives distances to 85 early-type galaxies in the cluster with relative accuracies of 5\% \citep{2007ApJ...655..144M, Blakeslee2009}.
The second is a product of the ground-based NGVS \citep{Cantiello2018, Cantiello2024}.  The earlier paper, limited to galaxies with $B_T \lesssim 13$, gives distances of 7\% relative accuracy to 77 cluster members with SBF \citep{Cantiello2018}.
There are 34 galaxies in common between this study and ACSVCS, resulting in distance measures for 128 objects (\citealt{Cantiello2024} also adds $\sim\,$200 more SBF distances, mainly with dwarf ellipticals).
The absolute distance calibration of the ACSVCS and the $B_T \lesssim 13$ NGVS surveys share the same origin: distances for 31 Virgo systems from the \citet{2002MNRAS.330..443B} revision of \citet{2001ApJ...546..681T}, itself established by Cepheid distances from the HST Key Project \citep{2001ApJ...553...47F}.
Because of this common calibration, distance moduli between the two studies for the 34 galaxies in common agree in the mean to within 0.001 mag, with scatter $\pm0.108$ mag.
Each of the two separate samples gives a mean SBF distance to the cluster of $16.5 \pm0.1$~Mpc (averaging the individual member distances, while excluding foreground interlopers, as well as background members of W$^{\prime}$), a value subsequently commonly taken as the distance of the Virgo Cluster.

Figure~\ref{fig:histDM} is a histogram of the distance moduli in the combined ACSVCS with HST and the NGVS from CFHT/MegaCam.  The mean and median values shown in the figure agree within 1\% in distance. 
The distance modulus values from these two earlier studies for the seven galaxies studied in this JWST program are shown in red.

\begin{deluxetable}{lcccc}
\tabletypesize{\small}
\tablewidth{0pt}
\tablehead{
\colhead{Target} & \colhead{Date} & \colhead{ID} & \colhead{Filter} & \colhead{Time [s]}}
\startdata
M\,49 (NGC 4472)      & 2024-06-09 & t006 & F090W & 5153.6496 \\
M\,49 (NGC 4472)      & 2024-06-09 & t006 & F150W & 1073.677 \\
M\,59 (NGC 4621)      & 2024-05-26 & t009 & F090W & 5153.6496 \\
M\,59 (NGC 4621)      & 2024-05-26 & t009 & F150W & 1073.677 \\
M\,60 (NGC 4649)      & 2024-06-26 & t010 & F090W & 5153.6496 \\
M\,60 (NGC 4649)      & 2024-06-27 & t010 & F150W & 1073.677 \\
M\,84 (NGC 4374)      & 2024-05-12 & t004 & F090W & 4294.708 \\
M\,84 (NGC 4374)      & 2024-05-12 & t004 & F150W & 1073.677 \\
M\,86 (NGC 4406)      & 2024-05-30 & t005 & F090W & 4294.708 \\
M\,86 (NGC 4406)      & 2024-05-30 & t005 & F150W & 1073.677 \\
M\,87 (NGC 4486)      & 2024-06-26 & t007 & F090W & 5153.6496 \\
M\,87 (NGC 4486)      & 2024-06-26 & t007 & F150W & 1073.677 \\
M\,89 (NGC 4552)      & 2024-06-26 & t008 & F090W & 5153.6496 \\
M\,89 (NGC 4552)      & 2024-06-26 & t008 & F150W & 1073.677 \\
M\,105 (NGC 3379)     & 2024-05-26 & t013 & F090W & 2576.8248 \\
M\,105 (NGC 3379)     & 2024-05-26 & t013 & F150W & 858.9416 \\
NGC 4636 & 2024-05-15 & t013 & F090W & 5153.6496 \\
NGC 4636 & 2024-05-15 & t013 & F150W & 1073.677 \\
NGC 4697 & 2024-06-26 & t013 & F090W & 2576.8248 \\
NGC 4697 & 2024-06-26 & t013 & F150W & 858.9416 \\
\enddata
\caption{Observation log for the JWST visits presented in this paper. The exposure times given are the values of \texttt{TMEASURE} from the image headers, which are the relevant times for S/N calculations. \label{tb:obs}}
\end{deluxetable}

\begin{figure*}
\epsscale{1.15}
\plotone{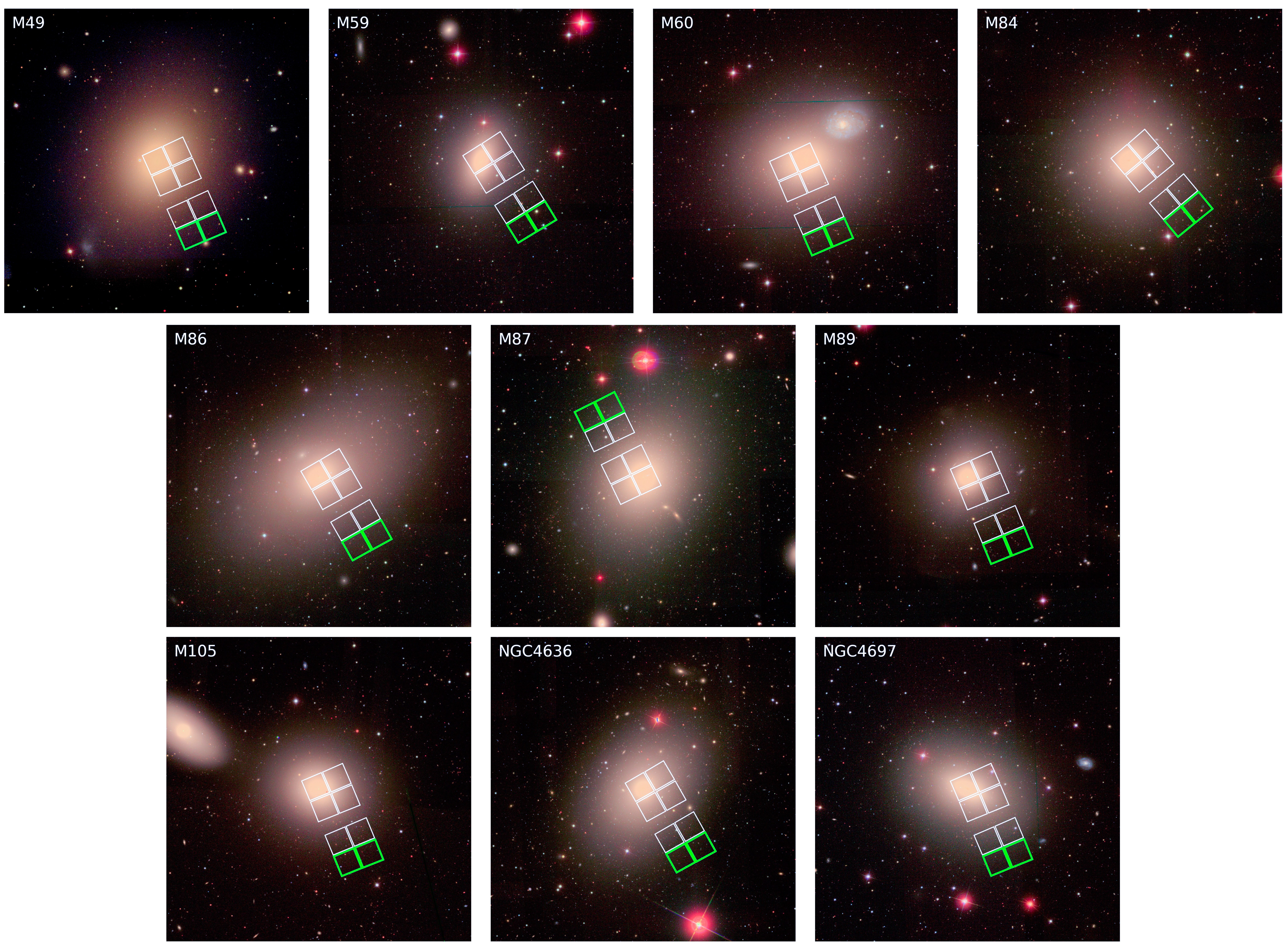}
\caption{Footprints of our JWST/NIRCam observations, overlaid on \SI{15}{\arcminute} $\times$ \SI{15}{\arcminute} \textit{gri} imaging from the Sloan Digital Sky Survey. NIRCam chips in green are those used for our TRGB analysis.}
\label{fig:footprints}
\end{figure*}

The left side of Figure~\ref{fig:allsbf} shows galaxies with Local Group velocities $< 3200$ \kms\ projected onto Virgo in equatorial coordinates.
The projected distribution of the 128 galaxies of the combined ACSVCS and NGVS samples is shown on the right side of Figure~\ref{fig:allsbf} along with the locations of the seven current JWST targets. This view is in supergalactic coordinates, a roughly $90^\circ$ clockwise rotation from the familiar equatorial presentation.   
The outer circle is centered on M87 and has a radius of $6.8^\circ$, roughly the expected radius of the second turnaround for a cluster mass of $6.3\times 10^{14}~\Msun$ \citep{2015AJ....149...54T}.
The distribution is flattened toward the supergalactic equator, but there are only mild concentrations toward substructures A, B and C around M87, M49 and M60, respectively. Although spatial patterns in systemic velocities are seen in the distribution of star forming galaxies, such as mentioned by \citet{Boselli2014}, no systematic flows are evident in the distribution of these predominantly old-population elliptical galaxies.
\citet{Cantiello2024} note an anomaly in the SBF distances of the sample extending to $B_T \sim 18$ in the vicinity of component A.  There is an apparent bi-modality with a dominant component about M87 in the foreground and $\sim 20\%$ of the sample including M84 and M86 lying $\sim 3$~Mpc to the background.


\section{Data and PSF Photometry} \label{sec:data}

\begin{figure*}
\epsscale{1.15}
\plotone{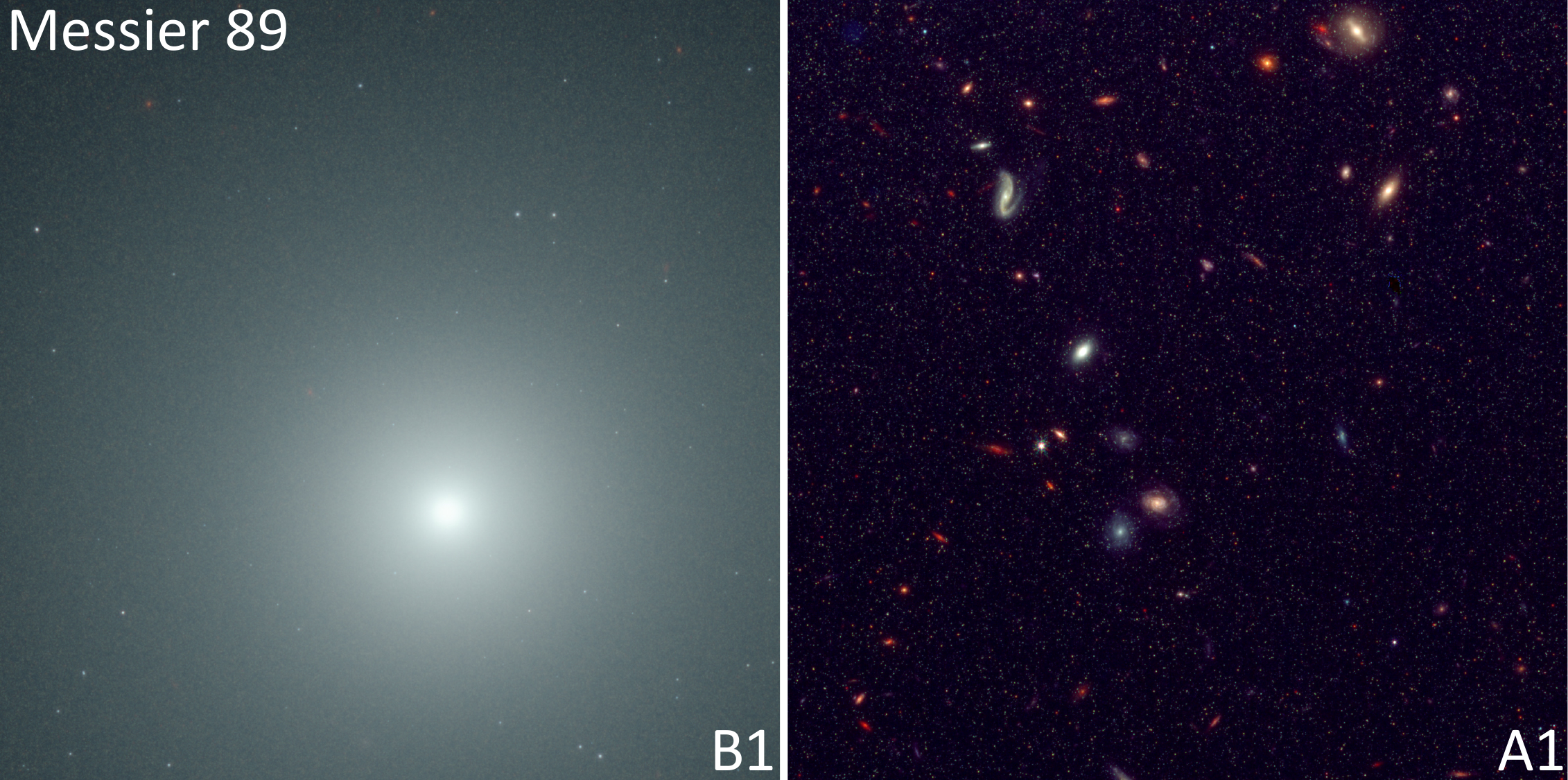}
\caption{$F090W$/$F150W$/$F356W$ color image of the core and halo of Messier~89. The extended spatial coverage of NIRCam allows us to obtain imaging that covers a very large dynamic range of surface brightness, facilitating both SBF (spatially unresolved) and TRGB (fully resolved) stellar population measurements. Pixels near the core of the galaxy were saturated in redder filters and were manually adjusted to create the color image.} 
\label{fig:m89color}
\end{figure*}

Table~\ref{tb:obs} provides a summary of the JWST/NIRCam data presented in this paper. The ten galaxies are those that are in and around the vicinity of the Virgo Cluster from the fourteen total galaxies observed as part of the Cycle~2 JWST program GO-3055~\citep{2023jwst.prop.3055T}. All the data were reduced with context versions (1230 $\leq$ \texttt{pmap} $\leq$ 1241) which include the latest updates to the photometric zero-points (provided in \texttt{1126.pmap}), with no significant differences relevant for our NIRCam photometry between these releases,\footnote{See \url{https://jwst-crds.stsci.edu/display_all_contexts/} for a description of each context version.} or the ones used in our Fornax work \citep{2024ApJ...973...83A}. Footprints of these NIRCam observations are shown in Fig.~\ref{fig:footprints}, overlaid on \SI{15}{\arcminute} $\times$ \SI{15}{\arcminute} \textit{gri} imaging from the Sloan Digital Sky Survey \citep{SDSS2015}. To showcase the range of surface brightness probed with a single NIRCam pointing, we show a color image of both the core and halo of Messier 89 in Figure \ref{fig:m89color}. The B1 chip contains the elliptical galaxy core, which is perfectly suited for SBF measurements. The color image of the A1 field shows the much lower surface brightness and less crowded region in the galaxy's halo which is perfectly suited to the TRGB measurement. For all of our targets, we use resolved star photometry obtained from the halo region (the outermost 2/8 of the NIRCam short-wavelength chips). Parallel NIRISS imaging is available for 11 of 14 of the pointings from GO-3055. We do not present results from NIRISS imaging in this work as we are currently unable to obtain highly precise PSF photometry due to the severe undersampling in $F090W$ at the NIRISS pixel scale (0.06$''$/pixel).

The resolved stellar photometry to measure the TRGB follows the same procedure described by \citet{2024ApJ...973...83A}. To briefly recap, we use the latest version (from 4 Feb 2024) of the DOLPHOT photometry suite \citep{2016ascl.soft08013D, Weisz2023} to perform PSF photometry on the individual stage 2 \texttt{cal} exposures while following the reduction parameters recommended by the JWST Resolved Stars ERS program \citep{Weisz2023, 2024ApJS..271...47W}. An initial estimate of the PSF is obtained with WebbPSF \citep{2014SPIE.9143E..3XP, 2015ascl.soft04007P} and available as part of the DOLPHOT package. Empirical adjustments to this starting PSF are made on a frame-by-frame basis, but are quite small, consistent with the findings of \cite{2024ApJS..271...47W}. Aperture corrections are performed on $\sim$200 bright, isolated stars per frame using the \texttt{ApCor} feature. Sources of poor quality are removed from the photometric catalogs via quality cuts. Specifically, only point sources that meet the following criteria are retained (adapted from \citealt{Warfield2023}): (1) Crowding $<$ 0.5; (2) $\mathrm{(Sharpness)^{2}}$ $\le$ 0.01; (3) Object Type $\le$ 2; (4) S/N $\ge$ 5; (5) Error Flag $\le$ 2. Our photometry is presented on the Vega$-$Vega zero-point system and not the Sirius$-$Vega variant.

\begin{figure*}
\epsscale{1.15}
\plotone{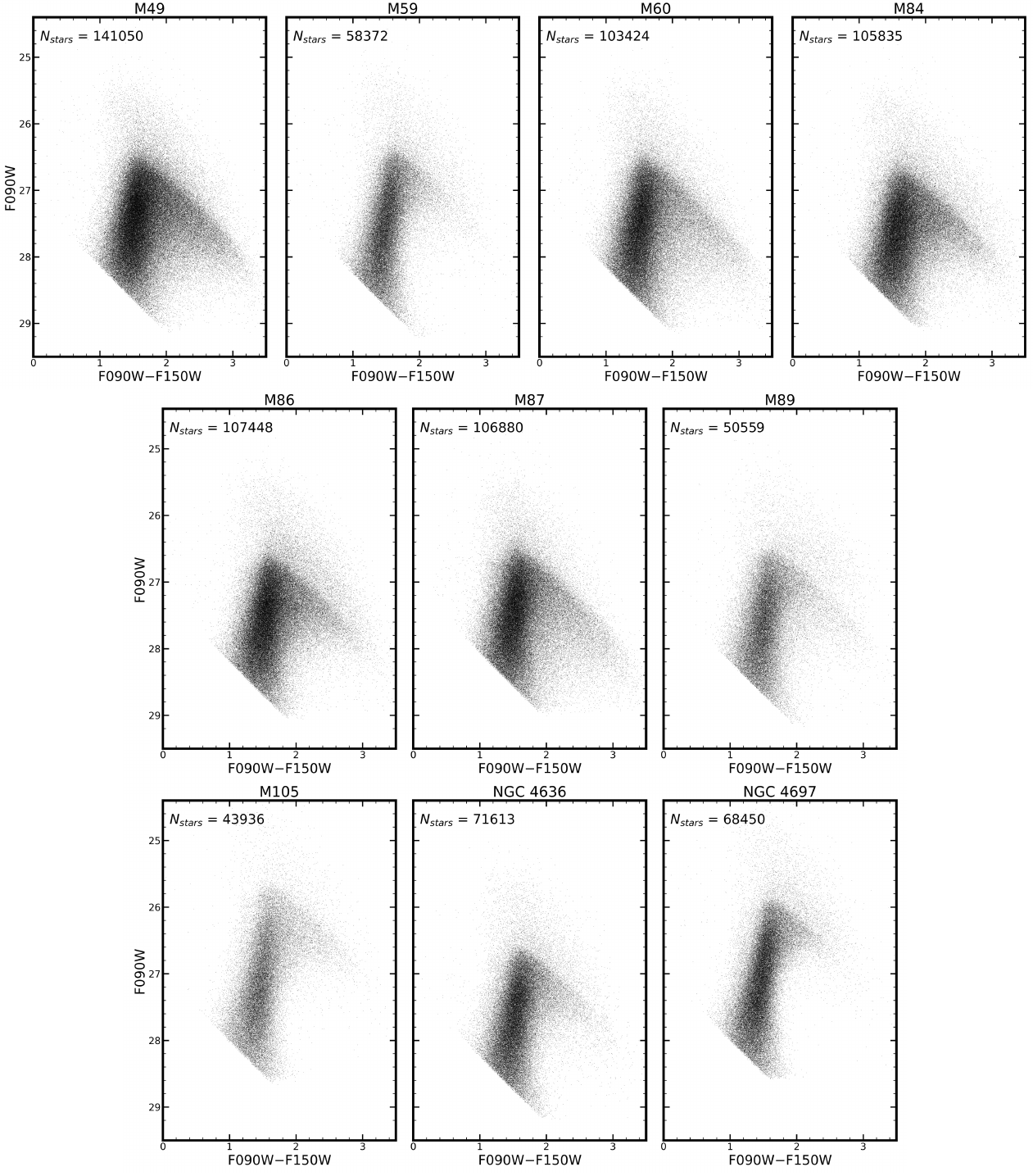}
2\caption{Color magnitude diagrams for each of the ten galaxies presented in this paper. The CMDs for each of the two chips are combined and displayed for each target, and the total number of stars is displayed in the top left. Only stars between 0.50 $\leq$ $F090W-F150W$ $\le$ 1.75~mag were used for the determination of the TRGB, due to the downward sloping of the TRGB at higher metallicities and redder colors 
\citep{2024ApJ...966...89A}.}
\label{fig:CMDs}
\end{figure*}

To measure the levels of photometric bias, errors, and incompleteness, we performed artificial star experiments using the same procedure as our sky star photometry. Utilizing DOLPHOT's artificial star mode, $\sim$200,000 stars are injected one at a time and recovered, with the same quality cuts applied to this test sample as were applied with the sky photometry. Color-magnitude diagrams (CMDs) for each of our galaxies can be seen in Fig.~\ref{fig:CMDs}.

\begin{figure}
    \centering
    \includegraphics[width=0.9\linewidth]{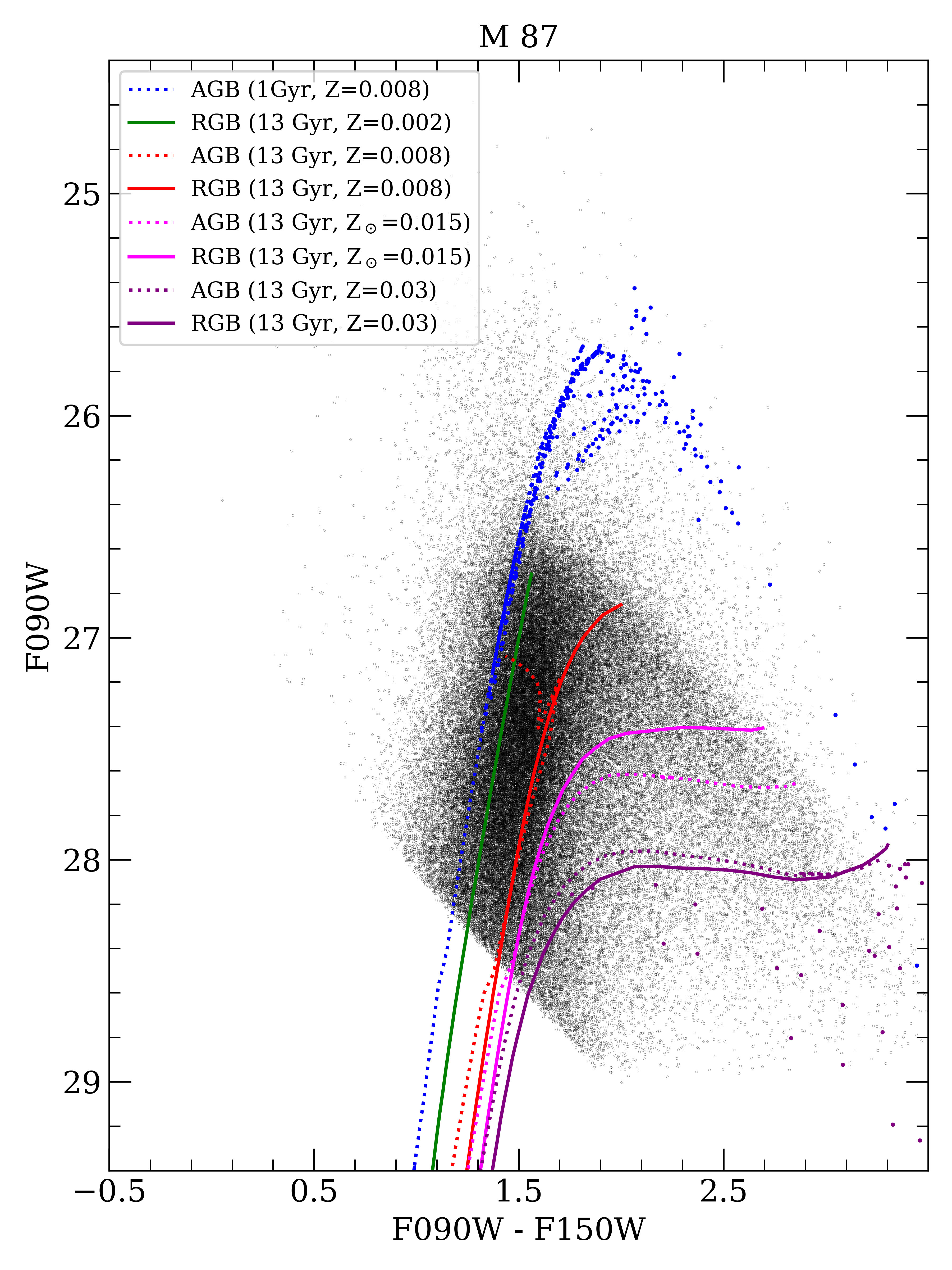}
    \caption{Color-magnitude diagram of M87 with the PARSEC isochrones overplotted. The isochrones were shifted according to the distances measured in this paper and Galactic extinction from \citet{2011ApJ...737..103S}. The models demonstrate that the downward slope of the TRGB is expected for uniformly old, increasingly metal-rich populations.}
    \label{fig:cmd_iso}
\end{figure}

Although at first look the ten CMDs appear similar, there are notable differences indicative of distinct metallicity spreads within each target. The $F090W-F150W$ color range is generally from 1.0$-$3.5 mag. The stars in M59, M105 and especially NGC4697 show a narrower range of colors, and hence metallicities. Intrinsically brighter galaxies (e.g. M49 and M87) show a more substantial tail toward redder colors, as expected from the wider metallicity spread foreseen in these more massive galaxies. There are practically no stars with a color less than 0.7 on our diagrams, indicative of a dearth of a young stellar population, in sharp contrast to the CMDs of nearby spirals in the same filters seen in other JWST programs \citep{2024ApJ...962L..17R, 2024arXiv240800065L}. Figure~\ref{fig:cmd_iso} shows the CMD of the central galaxy of the subcluster A, M87, with PARSEC isochrones \citep{2012MNRAS.427..127B,2013MNRAS.434..488M}. We see a wide variety of underlying metallicities for both AGB and RGB stars in this outer region of M87. Of particular note is that the CMDs typically reach 2.5$-$3.0 mags below the TRGB in $F090W$, deeper than our expectations from the JWST exposure time calculator (ETC) simulation runs for our proposal. Indeed, \cite{2024ApJ...970...36S} recently have described how to adjust the ETC to match the greater depths achievable with PSF photometry. 

\section{Distance Measurements} \label{sec:trgb}

\subsection{TRGB Measurements}

With the CMDs in hand, we proceed to measure the TRGB in the two of the eight NIRCam detectors which are the furthest away from the center of each galaxy (A1 and A2 in all cases, except B1 and B2 for M87). We chose to analyze each of these two fields separately due to the potential for detector-to-detector photometric offsets, which are currently expected to be at the level of up to 0.03~mag.\footnote{\url{https://jwst-docs.stsci.edu/jwst-calibration-status/nircam-calibration-status/nircam-imaging-calibration-status}} 

The measurement procedure follows the maximum-likelihood framework as described by \cite{2024ApJ...973...83A}, which uses the \texttt{TRGBTOOL} described in \cite{2006AJ....132.2729M}. In brief, the observed luminosity function of stars between colors of 0.50 $<$ $F090W-F150W$ $<$ 1.75 mag is modeled with a two-component power law, with the break denoting the magnitude of the TRGB. The functional form is given by

\begin{equation}
\psi = \begin{cases}
  10^{\,a(m-m_\mathrm{TRGB})+b}, & \mbox{if } m \geq m_\mathrm{TRGB} \\ 
  10^{\,c(m-m_\mathrm{TRGB})},   & \mbox{if } m < m_\mathrm{TRGB},
\end{cases}
\end{equation}
where $m_\mathrm{TRGB}$ is the magnitude of the TRGB, $b$ denotes the relative strength of the RGB discontinuity, and $a$ and $c$ are the slopes of the RGB and AGB, respectively. The procedure incorporates the observed photometric bias, errors, and completeness as determined from the artificial star experiments using Equations 4 and 5 in \cite{2006AJ....132.2729M}. 
For comparison purposes, we also ran Sobel filter measurements to provide an initial estimate of the TRGB magnitudes. 

The red end of the adopted color range is chosen to prevent the selection of TRGB stars of higher metallicities, which are fainter in magnitude in $F090W$ \citep{2024ApJ...966...89A,2024ApJ...973...83A}. While such stars could be used if the metallicity and age effects were understood, a complete calibration of these effects is the subject of a future study. The calibration provided by \cite{2024ApJ...966..175N,2024arXiv240603532N} does not extend out to the required redder colors needed to track the TRGB in the old, metal-rich populations of the elliptical galaxies in our program. The blue end of our selected color range is less critical, due to the lack of both 1) young stars that contaminate the luminosity function and 2) \textit{very} metal poor TRGB stars (M/H $<$ $-$2.0 dex, or $F090W-F150W$ $<$ 1.0 mag), which are expected to become fainter with magnitude \citep{2024ApJ...966...89A}.  We select a blue-side cutoff of $F090W-F150W$ = 0.50 mag to allow for a robust sampling of the underlying RGB luminosity function.

\begin{figure*}[p!]
\epsscale{1.15}
\plotone{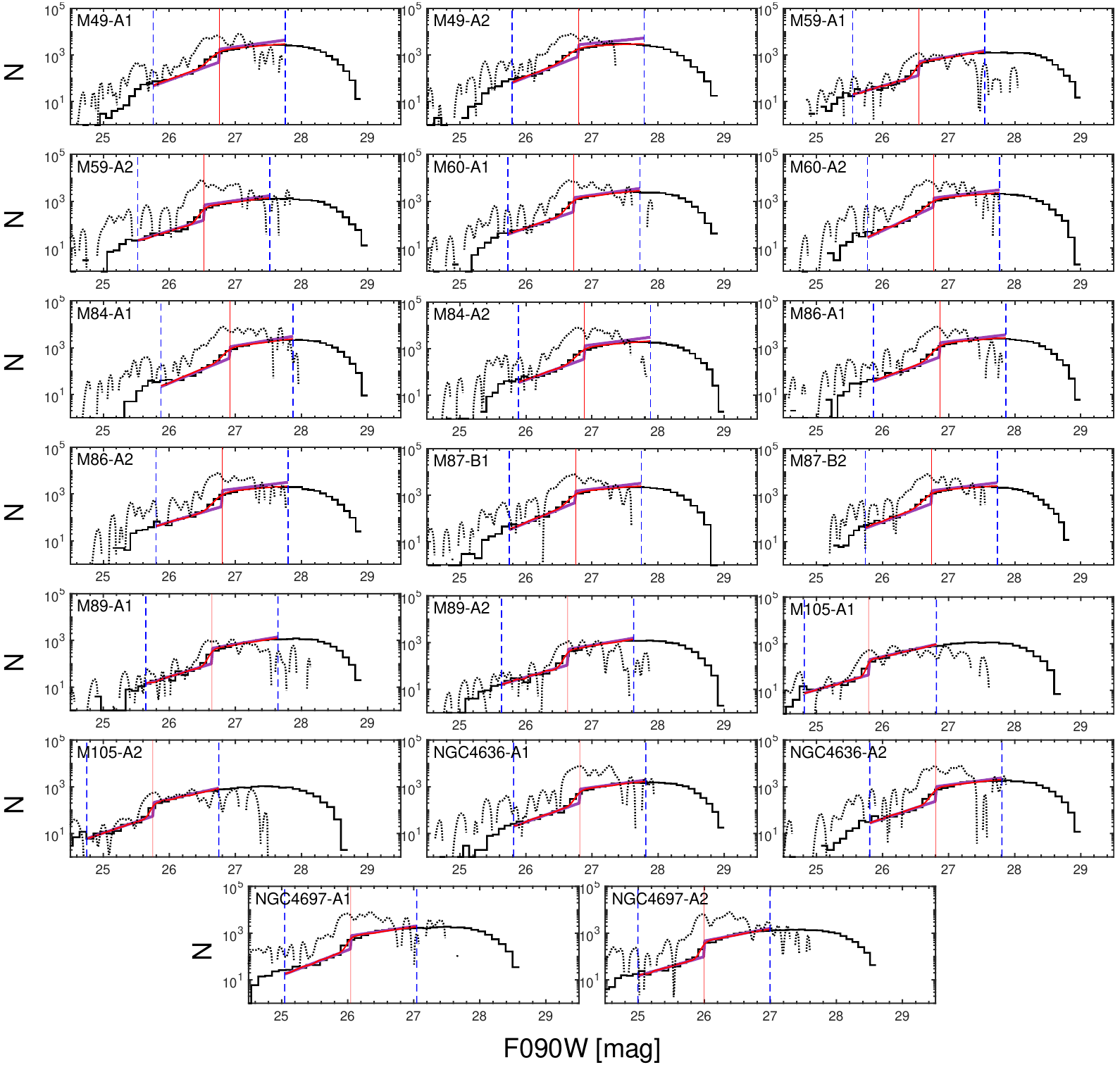}
\caption{Maximum-likelihood TRGB measurements for all 20 fields. A binned version of the luminosity function is shown as a histogram, purely for illustration (the measurement itself does not rely on binning). The magnitude of the TRGB is shown as the vertical red line, and the dashed blue lines denote the $\pm$1 magnitude range used for the fitting procedure. The best-fit luminosity function is shown as the bold red line, after accounting for the photometric bias, completeness, and errors as determined from our artificial star experiments with \texttt{DOLPHOT}. The intrinsic underlying luminosity function is shown in purple. Outputs from a Sobel edge-detection algorithm are shown with the dotted line for comparison.}
\label{fig:ml-example}
\end{figure*}

The TRGB measurements and their associated uncertainties for each of the 20 fields in 10 galaxies are provided in Table~\ref{tab:trgb-distances}. The maximum-likelihood fits themselves are shown in Figure \ref{fig:ml-example}. Besides the measurement error, additional statistical error terms arise from details in the TRGB color range and field selections (0.01 mag each), WebbPSF models/aperture corrections (0.02 mag), and NIRCam PSF stability (0.01 mag) as outlined in \cite{2024ApJ...966...89A}, as well as the uncertainty in the foreground extinction (where we conservatively adopt one-half of the extinction as an uncertainty floor). The uncertainties in the $m_{TRGB}$ values shown in Table~\ref{tab:trgb-distances} reflect the quadrature sum of these quantities.  

With 12 measured differences between the A1 and A2 detectors of NIRCam (9 from this paper, and 3 from \citealt{2024ApJ...973...83A}), there is a median difference of A1$-$A2 = 0.012~mag and a mean difference of A1$-$A2 = 0.010~mag. The error in the mean is 0.010~mag, with a standard deviation of 0.035~mag, consistent with no significant measurable differences between the chips. 

Results of a Sobel filter response applied to the Gaussian-smoothed luminosity function \citep{1996ApJ...461..713S} are shown in the same Figure \ref{fig:ml-example} as a dotted line, purely for comparison. The Sobel filter responses do not take into account the results of the artificial star experiments. \cite{2024ApJ...966...89A} and \cite{2024arXiv240800065L} both show JWST data where the Sobel responses are noisy, and in many cases multi-peaked, leading to ambiguity in which is the ``correct'' response to be associated with the TRGB. Regardless, with a reasonable selection of the Sobel filter output (selecting the brightest response within the vicinity of the appearance of the TRGB in the CMD), the measurements from the Sobel routine are on average 0.038~mag brighter (closer) with a standard deviation of 0.043~mag in the differences. The \cite{2024ApJ...966...89A} zero-point calibration for edge-detection routines ($-$4.377~mag) is correspondingly brighter by 0.030~mag compared to the maximum likelihood calibration ($-$4.347~mag), meaning that distances between the two methodologies, on average, would be consistent to within better than 0.01~mag. In general, the agreement between the maximum-likelihood and Sobel techniques is likely to be worse if the CMD below the TRGB is poorly sampled, as can be seen in Figure 9 of \cite{2006AJ....132.2729M}.

\begin{deluxetable*}{lcccccc}[t]
\tabletypesize{\small}
\tablewidth{0pt}
\tablehead{
\colhead{Galaxy}  & \colhead{$A_{F090W}$} & \colhead{$m_{TRGB}$ (A1)} & \colhead{$m_{TRGB}$ (A2)}  & \colhead{{$m_{50\%}$} (A1, A2)} & \colhead{$\mu$ [mag]} & \colhead{$D$ [Mpc]}}
\startdata
M49 (NGC 4472)  & 0.028 & 26.753 $\pm$ 0.032 & 26.790 $\pm$ 0.032 &  28.10, 27.80  & 31.091 $\pm$ 0.071 & 16.5 $\pm$ 0.5 \\ 
M59 (NGC 4621)  & 0.040 & 26.548 $\pm$ 0.038 & 26.519 $\pm$ 0.037 &  28.75, 28.57  & 30.841 $\pm$ 0.074 & 14.7 $\pm$ 0.5 \\ 
M60 (NGC 4649)  & 0.034 & 26.724 $\pm$ 0.034 & 26.771 $\pm$ 0.034 &  28.39, 28.38  & 31.061 $\pm$ 0.072 & 16.3 $\pm$ 0.5 \\ 
M84 (NGC 4374)  & 0.051 & 26.914 $\pm$ 0.040 & 26.883 $\pm$ 0.039 &  28.42, 28.25  & 31.195 $\pm$ 0.075 & 17.3 $\pm$ 0.6 \\ 
M86 (NGC 4406)  & 0.037 & 26.863 $\pm$ 0.034 & 26.795 $\pm$ 0.034 &  28.35, 28.19  & 31.139 $\pm$ 0.072 & 16.9 $\pm$ 0.6 \\ 
M87 (NGC 4486)  & 0.028 & 26.746 $\pm$ 0.033 & 26.740 $\pm$ 0.032 &  28.31, 28.18  & 31.062 $\pm$ 0.071 & 16.3 $\pm$ 0.5 \\ 
M89 (NGC 4552)  & 0.051 & 26.644 $\pm$ 0.040 & 26.630 $\pm$ 0.041 &  28.72, 28.65  & 30.933 $\pm$ 0.075 & 15.4 $\pm$ 0.5 \\ 
M105 (NGC 3379) & 0.031 & 25.789 $\pm$ 0.042 & 25.749 $\pm$ 0.040 &  28.33, 28.25  & 30.085 $\pm$ 0.076 & 10.4 $\pm$ 0.4 \\ 
NGC 4636        & 0.037 & 26.813 $\pm$ 0.035 & 26.806 $\pm$ 0.035 &  28.78, 28.63  & 31.120 $\pm$ 0.072 & 16.8 $\pm$ 0.6 \\ 
NGC 4697        & 0.037 & 26.045 $\pm$ 0.036 & 25.995 $\pm$ 0.036 &  28.19, 28.20  & 30.330 $\pm$ 0.073 & 11.6 $\pm$ 0.4 \\ 
\enddata 
\caption{Foreground extinction values, TRGB measurements for the two separate chips (A1 and A2 for all cases except M87, where they are B1 and B2) and their associated statistical uncertainties, 50$\%$ completeness limits in $F090W$ derived from our artificial star experiments, and averaged distance moduli (and corresponding distances) for each galaxy (including the uncertainty in the zero-point scaling).}
\label{tab:trgb-distances}
\end{deluxetable*}

\subsection{Distances}

To provide an absolute scaling for our TRGB measurements, we adopt the zero-point calibration of $M_{TRGB}^{F090W}$ = $-$4.347 $\pm$ 0.033~(stat) $\pm$~0.054~(sys)\footnote{\cite{2024ApJ...966...89A} provide a systematic error term of 0.045~mag, which was increased by \cite{2024ApJ...973...83A} to 0.054~mag.}~mag obtained via the identical maximum-likelihood procedure used in this work \citep{2024ApJ...966...89A,2024ApJ...973...83A}. This zero-point is derived from JWST/NIRCam observations within the outer regions of NGC~4258, which is host to a water megamaser, allowing for the determination of a highly precise geometrical distance \citep{2019ApJ...886L..27R}. The total uncertainty in this zero-point scaling becomes a systematic error term for the measurement of our distances, and is given by the quadrature sum of $\pm$0.063~mag. 

All of our targets suffer from a minor amount of foreground extinction, corrected by adopting the color excess from \citep{2011ApJ...737..103S} as retrieved through NED, while assuming $A_{F090W}$/[E(B$-$V)] = 1.4156 \citep{2024ApJ...966...89A}. After correcting for extinction and applying the zero-point calibration, we obtain averaged distance moduli, as shown in the last column of Table \ref{tab:trgb-distances}.

To obtain the overall uncertainties, we take the statistical uncertainty on the individual tip measurement as shown in Table \ref{tab:trgb-distances} and add in quadrature the uncertainty in the zero-point calibration ($\pm$0.063~mag) to arrive at the final value. For the reported distance modulus to each galaxy, we take the average of each of the two fields without further reduction of the underlying uncertainties. 

Of particular note is our derived distance to M87 of $\mu = 31.062 \pm 0.071$~mag, or $d = 16.3 \pm 0.5$~Mpc. M87 is host to the first supermassive black hole targeted by the Event Horizon Telescope (EHT) collaboration \citep{2019ApJ...875L...1E}. The EHT assumed a distance of $d = 16.8_{-0.7}^{+0.8}$~Mpc \citep{2019ApJ...875L...6E}, obtained by combining prior TRGB \citep{2010A&A...524A..71B} and SBF \citep{Blakeslee2009, Cantiello2018} distances. \cite{2010A&A...524A..71B} measured a TRGB distance modulus of $\mu$ = 31.12 $\pm$ 0.14~mag with extremely deep HST/ACS imaging ($\sim$27 hours) of M87, a value that is consistent with our own determination. 

\subsection{Distance to the Virgo Cluster}

How well do the seven JWST Virgo cluster distances represent those of the cluster as a whole? In this section we examine the relative JWST measurements and an alternative sample of SBF distances acquired with HST ACS imaging \citep{2007ApJ...655..144M, Blakeslee2009}.  The black filled histogram seen in Fig.\,\ref{fig:histDM} shows the distribution of 85 distance moduli for galaxies associated with the Virgo Cluster (and five galaxies in the background Virgo\,W$^{\prime}$ group). The red components of the histogram represent the placement of the seven JWST targets within the cluster {\it at the distances given by the HST ACS SBF program.}

The HST ACSVCS SBF sample \citep{2007ApJ...655..144M} is substantial (85 galaxies) and particularly useful because the galaxies are all early type (E and S0), ``red and dead'' populations appropriate for surface brightness fluctuation analysis.  By and large, it is expected that these systems formed or merged very early in the cluster and are well mixed. The distribution of galaxies seen in 2-dimensions in Fig.~\ref{fig:allsbf} and inferred 3-dimensions from SBF distances is consistent with this expectation. Spiral or gas-rich galaxies are found predominantly on the outskirts of the cluster and may have spatial distributions that poorly represent the cluster's mean gravitational potential.

The mean distance modulus for the 85 Virgo Cluster galaxies recorded by \citet{Blakeslee2009} is $31.092\pm 0.013$ mag and RMS variance 0.124 mag (5.9\% in distance).  The variance is an anticipated consequence of the depth of the cluster.  The distance of the cluster on the zero point scale of the SBF study is 16.5~Mpc, a value that has often been used as the Virgo Cluster distance in recent publications.

\begin{deluxetable*}{cccc}[t]
\tabletypesize{\small}
\tablewidth{0pt}
\tablehead{
\colhead{Galaxy} & \colhead{JWST TRGB} & \colhead{HST SBF} & $\Delta$ (TRGB$-$SBF)} 
\startdata
M49 (NGC 4472) &   31.091 $\pm$ 0.032  &  31.116 $\pm$ 0.075 & $-$0.025 $\pm$ 0.081 \\
M59 (NGC 4621) &   30.841 $\pm$ 0.038  &  30.859 $\pm$ 0.072 & $-$0.018 $\pm$ 0.081 \\
M60 (NGC 4649) &   31.061 $\pm$ 0.034  &  31.082 $\pm$ 0.079 & $-$0.021 $\pm$ 0.086 \\
M84 (NGC 4374) &   31.195 $\pm$ 0.040  &  31.337 $\pm$ 0.070 & $-$0.142 $\pm$ 0.081 \\
M86 (NGC 4406) &   31.139 $\pm$ 0.034  &  31.261 $\pm$ 0.072 & $-$0.122 $\pm$ 0.080 \\
M87 (NGC 4486) &   31.062 $\pm$ 0.033  &  31.111 $\pm$ 0.076 & $-$0.049 $\pm$ 0.083 \\
M89 (NGC 4552) &   30.933 $\pm$ 0.040  &  31.021 $\pm$ 0.073 & $-$0.088 $\pm$ 0.083 \\
\hline    
\hline     
JWST TRGB $-$ HST SBF  & $-0.067 \pm 0.031$ & RMS $\pm 0.051$  \\
Virgo (85 HST SBF) & $31.092 \pm 0.013$ & RMS $\pm 0.124$ & d = 16.53 Mpc \\
Virgo (TRGB recalib.) & $31.043 \pm 0.034$ & & d = 16.17 Mpc \\
\hline
\enddata 
\caption{A comparison of distance moduli to the Virgo Cluster galaxies, from JWST TRGB (this work) and HST SBF \citep{2007ApJ...655..144M} measurements, where the uncertainties listed here \textit{exclude} correlated (systematic) uncertainties for each individual method. We determine mean and median offsets of $\Delta \mu_{TRGB-SBF}$ = $-$0.067~mag and $-$0.049~mag respectively, and use the median offset to derive a new distance to the Virgo Cluster of d $=$ 16.17 $\pm$ 0.25 (stat) $\pm$ 0.47 (sys) Mpc.}
\label{tab:qual-distances}
\end{deluxetable*}

The seven JWST TRGB distance measurements are too few by themselves to give a reliable distance to the cluster.  However, it is informative to compare these new TRGB distances with the HST SBF distances for the same targets.  The comparisons are compiled in Table~\ref{tab:qual-distances}.  The distance moduli from the TRGB measurements are consistently lower than those of the HST SBF scale. The weighted mean difference, using the error analysis methods outlined in \cite{1997ieas.book.....T} and weights of 1/$\sigma^{2}$, is $\Delta \mu_{TRGB-SBF} = -0.067$ $\pm$ $0.031$~mag with RMS scatter $\pm0.051$~mag. This scatter of 2.4\% in distance is below the 3 to 3.5\% individual uncertainties given in Table~\ref{tab:qual-distances} for the SBF measurements. The mean offset of $-$0.067 mag in the modulus has $2.2 \sigma$ significance. However, we note that the two galaxies with the highest significance in their individual offsets are M84 and M86, which lie behind the core of the cluster. It is possible that these larger offsets are driven by the TRGB measurements being influenced by a small number of intracluster stars \citep{2005ApJ...631L..41M, 2007ApJ...656..756W} lying to the foreground of these galaxies. 
If present, intracluster foreground contamination would bias the TRGB measurement to be slightly brighter (nearer distance), while the SBF measurements at high surface brightness would not be affected by such contamination.
In particular, \cite{2007ApJ...656..756W} note that the onset of the TRGB in their intracluster field was not as sharp as expected, consistent with a cluster depth of a few megaparsecs along the line of sight. We examined the parameter $b$ which denotes the relative strength of the RGB discontinuity (see Equation 1), and can serve as a direct probe of this effect. On average, our fields have $b =$ 0.585 with a standard deviation of $\pm$0.076~mag. The M84 and M86 fields have $b =$ 0.527 and $b =$ 0.653, respectively, not indicative of significant broadening of the luminosity function due to intracluster stars. Still, with a median value being more robust to potential outliers, we adopt an offset of $\mu_{TRGB-SBF} = -0.049$~mag. 

Applying the adjustment of $-0.049$~mag to the ensemble of HST bf ACSVCS SBF measurements ($31.092$~mag) gives the cluster distance modulus $31.043$ $\pm$ 0.034 (stat) $\pm$ 0.063 (sys) mag, where the statistical term is the quadrature sum of 1) the uncertainty in the tabulation of averages of differences between TRGB and SBF distance moduli and 2) the uncertainty in the HST SBF distance moduli for 85 cases. The systematic term arises from the uncertainty in the adopted TRGB zero-point. The corresponding distance is $d=16.17$ $\pm$ 0.25 (stat) $\pm$ 0.47 (sys)~Mpc.

\begin{figure}
\epsscale{1.23}
\plotone{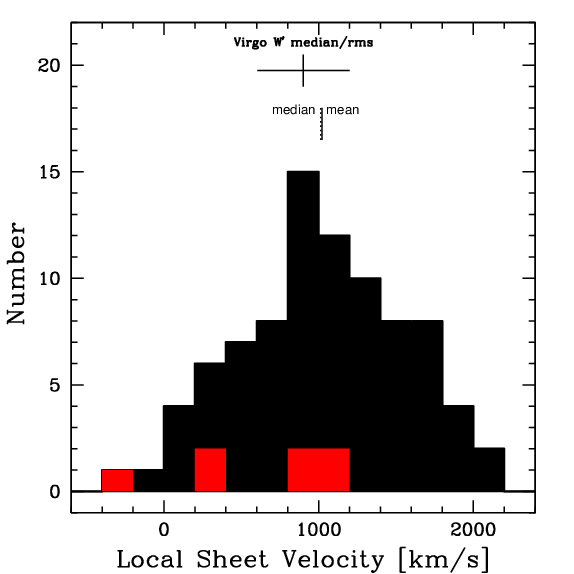}
\caption{Histogram of velocities \citep{2009AJ....138..323T, Kourkchi2017} of the 85 galaxies in the HST SBF sample in black and the 7 galaxies in the JWST TRGB sample in red.  The mean and median velocities are almost the same: 1023~\kms and 1013~\kms, respectively, in the Local Sheet reference frame. The mean and variance of velocities in the Virgo~W$^{\prime}$ group are shown at the top of the figure.}
\label{fig:histV}
\end{figure}

For completeness, Fig.~\ref{fig:histV} shows the distribution of systemic velocities for the HST SBF sample-- velocities are available at the Extragalactic Distance Database \citep{2009AJ....138..323T} in the file {\it Kourkchi-Tully Groups} \citep{Kourkchi2017} associated with the Virgo Cluster = 1PGC\,43296.  The Local Sheet velocity reference frame is a Local Group variant \citep{2008ApJ...676..184T}.  In that reference frame, the mean velocity for the 85 galaxies in the HST SBF sample is $1023 \pm 58$~\kms\ with RMS scatter $\pm535$~\kms. This scatter is less than for the members of the cluster as a whole, commensurate with the expectation that the old E$-$S0 galaxies of the HST SBF sample are concentrated in the center of the cluster potential. The mean heliocentric velocity of the sample is 1115~\kms, or 1444~\kms\ in the frame of the cosmic microwave background. 
The Virgo Cluster is so near that peculiar velocities invalidate any inference of $H_0$ from direct measurements of the cluster distance and velocity.


\subsection{More on Cluster Substructure}

In Section~2, we mentioned substructures in the distribution of early-type galaxies, which some break into components A, B, C \citep{deVaucouleurs1961, Boselli2014}.
The two dominant galaxies in the cluster, M87 in the central substructure A and M49 in the peripheral substructure B, are found to have distances of 16.3 and 16.5 Mpc, respectively (using the NGC\,4258 maser-derived zero point for TRGB, as with all the measurements in this study). The difference of 0.2 Mpc (1.2$\%$) in relative distances is not statistically significant, given the individual TRGB magnitude measurements are uncertain to within $\sim$1.5$\%$ (discounting the shared uncertainty in the NGC~4258-derived zeropoint). Likewise, the dominant galaxy in substructure C, M60 at 16.3~Mpc, is at the same distance as the two dominant galaxies within the uncertainties.
Two of the JWST targets lie somewhat in the foreground in the cluster: M59 lies at 14.7~Mpc in substructure C, and M89 lies at 15.4~Mpc in substructure A.  The other two JWST targets, associated with substructure A, lie towards the background: M84 is at 17.3~Mpc and M86 is at 16.9~Mpc. 
These offsets are in the same sense, although not quite as large, as found by \citet{Cantiello2024}.
The locations of M84 and M86 affirm the prescient prediction by \citet{Binggeli1993}, based only on velocities, that these systems are falling toward M87 from behind.

\subsection{Virgo Infall}

\begin{figure}
\epsscale{1.1}
\plotone{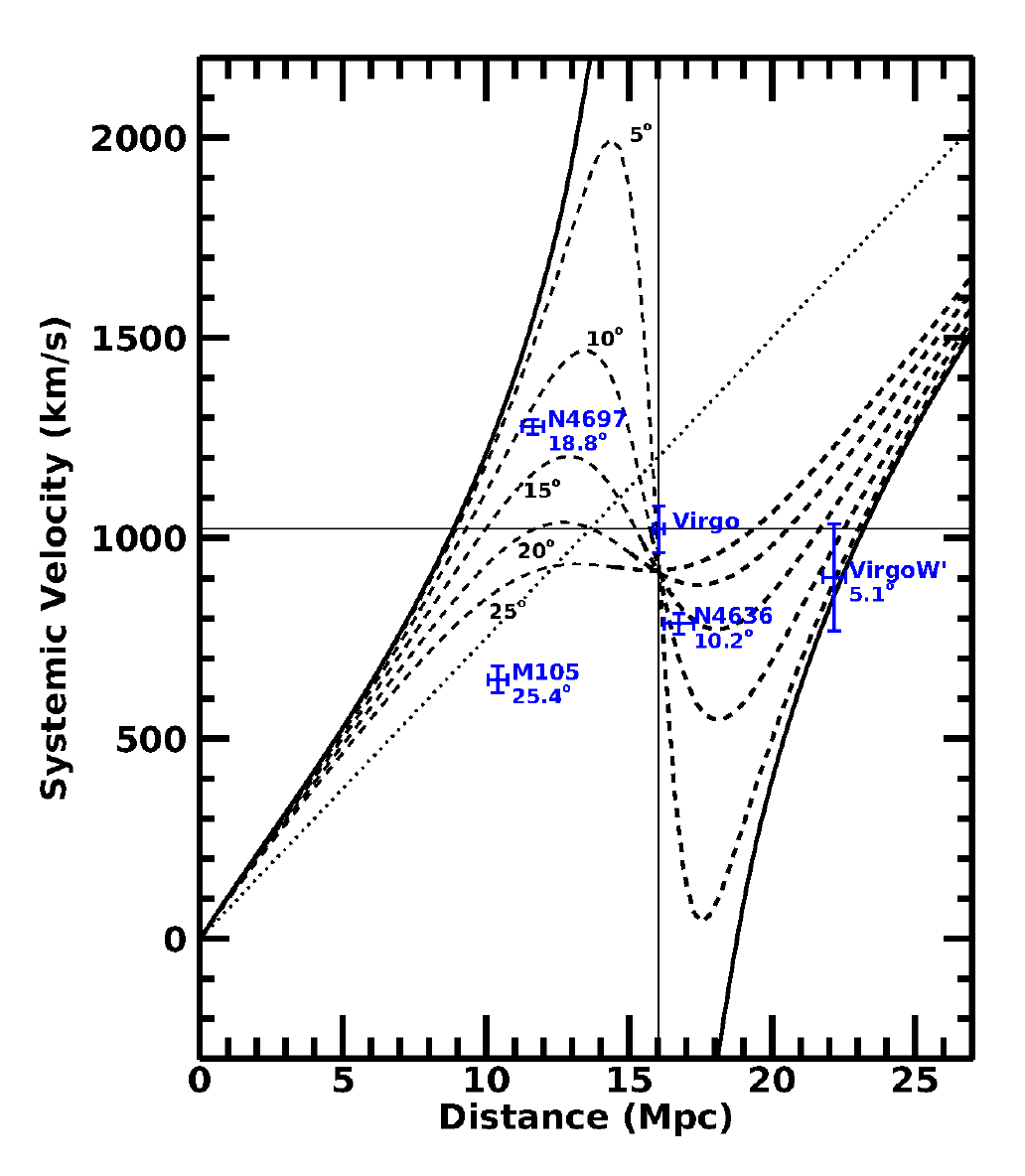}
\caption{
Distances and velocities of systems near the Virgo Cluster discussed in the text superimposed on a model Virgo infall pattern. The name and angular separation from M87 is provided alongside each distance-velocity point.  The solid line shows the pattern in the Virgo Cluster direction, and the dashed triple-value curves depict the velocity pattern along lines-of-site with angular separations from M87 as labeled on each curve. The diagonal dotted line is for a pure Hubble flow centered on the Local Group.} 
\label{fig:infall}
\end{figure}

The present JWST program includes three galaxies within or at the edge of the Virgo Cluster infall region: NGC\,4636, NGC\,4697, and M105 (NGC\,3379).
Accurate measurements of distances to galaxies in close proximity to the cluster provide constraints on the mass of the cluster \citep{hoffman1982, Tully1984, Kashibadze2020}.
Orbit reconstructions with numerical action methods that address three-dimensional complexity and the role of external tidal influences provide a mass estimate of the cluster as well as of the mass as a function of radius beyond the cluster \citep{2017ApJ...850..207S, Shaya2022}.
The number of new observations is insufficient to warrant a detailed re-analysis, but that there is consistency with the earlier work is demonstrated in Fig.~\ref{fig:infall}.
The model represented in this figure is a spherically symmetric model with a central mass of $6\times 10^{14}~\Msun$ increasing to $8.3 \times 10^{14}~\Msun$ at 7~Mpc from the cluster center and to $13.8 \times 10^{14}~\Msun$ in a sphere out to the Milky Way 
in a universe with $\Omega_m = 0.244$, $\Omega_\Lambda = 1 - \Omega_m$, and $H_0 = 75$ km\,s$^{-1}$\,Mpc$^{-1}$ (the value of $\Omega_m$ was chosen to be consistent with the higher values of $H_0$ measured in the local universe and is not important for the present discussion). 
An observation consistent with the model would lie along a triple-value curve \citep{1981ApJ...246..680T} commensurate with the angular separation of the object from the center of the cluster. For any given observed velocity, there are three prospective distance locations: departing or returning to Hubble flow or in close proximity to the cluster.
Curves in the figure show model loci at 5-degree intervals out to $25^{\circ}$ radius about the cluster.

Another set of observations merits consideration in this discussion.  The Virgo\,W$^{\prime}$ Group projects fully onto the Virgo Cluster.  The clear distance offset of 0.7~mag to the background based on HST SBF measurements \citep{2007ApJ...655..144M, Blakeslee2009} is seen in Fig.~\ref{fig:histDM}.
The galaxies of Virgo~W$^{\prime}$ lie near the backside location of decoupling between cluster infall and cosmic expansion.

We now turn our attention to the three galaxies in our sample neighboring the cluster.  NGC\,4636 is of particular interest because of its immediate proximity to the cluster at the intermediate triple-value location.  As expected from its negative velocity with respect to the cluster mean, it is observed to be behind the cluster and moving toward us relative to the cluster.  The position information is ambiguous for modeling purposes because the infall trajectory lies close to the plane of the sky (nominally within $5^{\circ}$).
In any event, this galaxy and its associated group with 29 known members \citep{Kourkchi2017} will be within the splashback confines of the cluster within 500 million years.

In the numerical action orbit modeling, NGC\,4697 had been taken to be at the intermediate triple-value location, but the JWST observation places it unambiguously at the front position.  Its velocity placement in Fig.~\ref{fig:infall} is based on its associated group velocity although its individual velocity of 1080~\kms\ is very close to the model expectation.  At its angular distance of $19^{\circ}$ from the cluster core it can be seen from Fig.~\ref{fig:infall} that galaxies ranging from 11 to 21 Mpc can have very similar observed velocities.

M105 (NGC\,3379) in the nearby Leo Group appears to have a velocity significantly below the model, but its location was anticipated by the numerical action orbit reconstructions \citep{2017ApJ...850..207S, Shaya2022}.  Those studies made clear that all nearby galaxies out to the Virgo Cluster and beyond that lie near the supergalactic equator are being pushed with velocities of ${\sim}200$~\kms\ toward the negative supergalactic pole by the expansion of the Local Void \citep{2019ApJ...880...24T}. 
M105 and its associated Leo Group lie in the separate Leo Spur structure below the equatorial band and are not experiencing this motion away from the Local Void.
The offset of M105 illustrates the need for a three-dimensional analysis that properly accounts for the complexities such as carried out by \citet{2017ApJ...850..207S, Shaya2022}.

\section{On the Zero-Point Calibration of the TRGB}

The absolute zero-point scaling for our program originates with the geometrical maser distance to NGC~4258 \citep{2019ApJ...886L..27R}, which was then used to determine the absolute magnitude of the TRGB in F090W with the measurement methodology described by \cite{2024ApJ...966...89A} and \cite{2024ApJ...973...83A}, giving $M_{TRGB}^{F090W}$ = $-$4.347 $\pm$ 0.033~(stat) $\pm$~0.054~(sys) mag in the Vega-Vega photometric system.

At this relatively early stage in the lifespan of JWST, there is only one other absolute calibration available in the $F090W$ filter: \cite{2024arXiv240603532N} report a value of $M_{TRGB}^{F090W}$ = $-$4.32 $\pm$ 0.025~(stat) $\pm$~0.043~(sys) mag in the Sirius-Vega photometric system (M. Newman, priv. comm.), which becomes  $M_{TRGB}^{F090W}$ = $-$4.36 $\pm$ 0.025~(stat) $\pm$~0.043~(sys) mag in our preferred choice of the Vega-Vega photometric system. The details of their calibration are outlined in \cite{2024ApJ...966..175N,2024arXiv240603532N}, and involve tracing the JWST calibration in a path through the HST photometric system. Of important note is that their underlying zero-point is set to that determined by the CCHP \citep{2021ApJ...919...16F}, which has multiple anchors in Milky Way globular clusters \citep{2020arXiv201209701C}, the LMC \citep{2021arXiv210613337H}, SMC \citep{2021arXiv210613337H}, and NGC~4258 \citep{2021ApJ...906..125J}. It is encouraging that the JWST $F090W$ calibration by \cite{2024arXiv240603532N} agrees with our present result to within better than 0.02~mag, or 1$\%$ in distance.

An alternative calibration pathway to the previously described geometric anchors would involve the utilization of Gaia parallaxes \citep{2016A&A...595A...1G}. A goal of one JWST Cycle 3 program \citep{2024jwst.prop.4783S} is to establish a JWST population II distance scale through multiple population II distance indicators, including the horizontal branch and the TRGB. Other efforts involve the calibration of RR Lyrae luminosities through Gaia \citep{2022ApJ...932...19N, 2022ApJ...938..101S}, leading to calibrations in the JWST filter system \citep{2024AAS...24410109S}. 

Our team is currently at work performing a color calibration of the TRGB in F090W that will extend to the reddest colors seen in our CMDs. The calibration provided by \cite{2024arXiv240603532N} determined a slope consistent with zero between 1.15 $ < F090W - F150W < $ 1.68~mag, but they lack a sufficient sampling of stars outside of this somewhat narrow color range. As can be seen in Figure \ref{fig:CMDs}, the TRGB slope for high metallicity stars is steep (and hence our exclusion of these stars from the present analysis). For future work involving the full range of the CMD, it will be crucial to take into account this metallicity (and hence color) dependence. The giant elliptical galaxies in the nearby Virgo and Fornax clusters observed with JWST will allow us to calibrate the color-magnitude relation of the TRGB over a wide range of colors. 

Our current absolute calibration rests on a single geometrical anchor---the maser distance to NGC\,4258. The establishment of additional geometrical calibration pathways will be required as we approach the ultimate goal of a 1$\%$ measurement of the Hubble constant based on Population II stars alone.


\section{Summary and Future Outlook} \label{sec:conc}

JWST observations from a Cycle 2 program \citep{2023jwst.prop.3055T} of ten galaxies in and around the Virgo Cluster allow us to measure highly precise TRGB distances to these targets. With these distances, we are able to resolve some of the substructure within the Virgo cluster and provide constraints on Virgo infall. After comparing distances to seven Virgo Cluster galaxies in common between this program and the SBF distances from the ACS Virgo Cluster Survey, we provide a re-calibrated distance to Virgo of $d = 16.17$ $\pm$ 0.25 (stat) $\pm$ 0.47 (sys)~Mpc.

When combined with TRGB distances to four more targets from this program and future SBF measurements of the same galaxies with the same dataset, the results of our program will provide a secure second-rung of a Population II distance ladder. The next step of our program will be to calibrate the color-dependency of SBF magnitudes with observations of ${\sim}50$ massive elliptical galaxies in the Coma Cluster as part of an approved Cycle 3 program \citep{2024jwst.prop.5989J}. Additional work is also required on the first rung of our distance ladder, as we are currently limited to a single geometric anchor (NGC~4258) on the JWST filter system. A full zero-point calibration of the TRGB with color will allow us to minimize potential systematics as a function of metallicity and age, while retaining the full breadth of the RGBs visible in our CMDs.

Lastly, SBF observations out into the Hubble flow with JWST/NIRCam would allow for the measurement of the Hubble constant with the same instrumental systematics from end-to-end, minus the initial geometric scaling. This would be an incredibly valuable result for the community at large and would set the foundation for an independent measurement of the Hubble constant that would be competitive with the results from Cepheids and Type Ia supernovae \citep{2022ApJ...934L...7R}. 


\begin{acknowledgments}
This work is based on observations made with the NASA/ESA/CSA JWST. The data were obtained from the Mikulski Archive for Space Telescopes at the Space Telescope Science Institute, which is operated by the Association of Universities for Research in Astronomy, Inc., under NASA contract NAS 5-03127 for JWST. These observations are associated with program \#3055. The specific observations analyzed can be accessed via \href{https://archive.stsci.edu/doi/resolve/resolve.html?doi=10.17909/9npf-6b57}{DOI 10.17909/9npf-6b57}.

G.S.A.{}, Y.C.{}, R.B.T. and J.B.J.\ acknowledge financial support from JWST GO–3055.
D.I.M.{} , L.N.M.{} and M.I.C.{} are supported by the grant \textnumero~075--15--2022--262 (13.MNPMU.21.0003) of the Ministry of Science and Higher Education of the Russian Federation. MC acknowledges support from the project ``INAF-EDGE" (Large Grant 12-2022, P.I. L. Hunt), and by Italian MIUR grant PRIN 2022 2022383WFT ``SUNRISE”, CUP C53D23000850006.

This research made use of the NASA Astrophysics Data System. This research also used the NASA/IPAC Extragalactic Database (NED), which is funded by the National Aeronautics and Space Administration and operated by the California Institute of Technology.

\end{acknowledgments}

\bibliography{paper}{}
\bibliographystyle{aasjournal}

\end{document}